\documentclass[twocolumn,letter]{jpsj2} 
\title{Evolution of an Unconventional Superconducting State inside the Antiferromagnetic Phase of CeNiGe$_3$ under Pressure: a $^{73}$Ge-Nuclear-Quadrupole-Resonance Study}
\author{Atsushi \textsc{Harada}$^{1}$\thanks{E-mail address: HARADA.Atsushi@nims.go.jp}, Hidekazu \textsc{Mukuda}$^{1}$, Yoshio \textsc{Kitaoka}$^{1}$, Arumugamu \textsc{Thamizhavel}$^{2}$, Yusuke \textsc{Okuda}$^{2}$,\\ Rikio \textsc{Settai}$^{2}$, Yoshichika \textsc{\=Onuki}$^{2}$, Kohei M. \textsc{Itoh}$^{3}$, Eugene \textsc{Haller}$^{4}$, and Hisatomo \textsc{Harima}$^{5}$}

\inst{$^{1}$Department of Materials Engineering Science, Osaka University, Osaka 560-8531, Japan \\
$^{2}$Department of Physics, Osaka University, Osaka 560-0043, Japan \\
$^{3}$Department of Applied Physics and Physico-Informatics,
Keio University, Yokohama 223-8522, Japan\\
$^{4}$Department of Materials Science and Engineering, University of California at Berkeley and Lawrence Berkeley National Laboratory, Berkeley, CA 94720, USA\\
$^{5}$Department of Physics, Faculty of Science, Kobe University, Nada, Kobe 657-8501, Japan}

\recdate{\today}

\abst{We report a $^{73}$Ge nuclear-quadrupole-resonance (NQR) study on novel evolution of unconventional superconductivity in antiferromagnetic (AFM) CeNiGe$_3$. The measurements of the $^{73}$Ge-NQR spectrum and the nuclear spin-lattice relaxation rate ($1/T_1$) have revealed that the unconventional superconductivity evolves inside a commensurate AFM phase around the pressure ($P$) where N\'{e}el temperature $T_{\rm N}$ exhibits its maximum at 8.5\,K. The superconducting transition temperature $T_{\rm SC}$ has been found to be enhanced with increasing $T_{\rm N}$, before reaching the quantum critical point at which the AFM order collapses. Above $T_{\rm SC}$, the AFM structure transits from an incommensurate spin-density-wave order to a commensurate AFM order at  $T\sim 2$\,K, accompanied by a longitudinal spin-density fluctuation. With regard to heavy-fermion compounds, these novel phenomena have hitherto never been reported in the $P$-$T$ phase diagram.}

\kword{heavy fermion, superconductivity, CeNiGe$_3$, commensurate and incommensurate antiferromagnetism, NQR under pressure}

\newpage
\begin{document}
\maketitle


Since the discovery of the heavy-fermion (HF) superconductor CeCu$_2$Si$_2$ \cite{Steglich}, the interplay between superconductivity and magnetism has been one of the most attractive subjects in condensed matter physics. The HF superconductivity often appears near a quantum critical point (QCP) where antiferromagnetism is suppressed by an application of pressure ($P$) in cerium(Ce)-based compounds such as CeCu$_2$Ge$_2$ \cite{Jaccard}, CePd$_2$Si$_2$ \cite{Grosche}, CeRh$_2$Si$_2$ \cite{Movshovich}, CeIn$_3$ \cite{Mathur}, and CeRhIn$_5$ \cite{Hegger}. Because of strong antiferromagnetic (AFM) correlation near the QCP, the above finding suggests that the mechanism forming Cooper pairs can be magnetic in origin. Namely, near the QCP, the magnetically soft electron liquid can mediate spin-dependent attractive interactions between the charge carriers.  In addition to the emergence of $P$-induced superconductivity above the QCP, another interesting phenomenon is the coexistence of superconductivity and antiferromagnetism below the QCP \cite{Mito,Yashima1,Kawasaki1}. 

\begin{figure}[h]
\centering
\includegraphics[width=7.0cm]{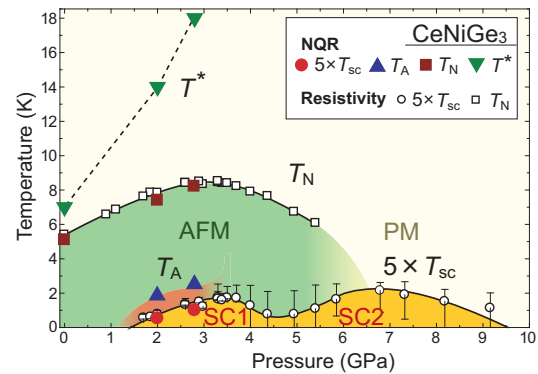}
\caption[]{(color online) Pressure~vs.~temperature phase diagram of CeNiGe$_3$ established by resistivity \cite{Kotegawa} along with the results of this study. The error bars represent the width of the SC transition between its onset and zeroresistivity, respectively \cite{Kotegawa}. We revealed a transition of the AFM structure, denoted as $T_{\rm A}$ at $P=2.0$ and 2.8\,GPa where SC1 takes place (see text). }
\label{fig1}
\end{figure}

The HF antiferromagnet CeNiGe$_3$ ($T_{\rm N}=5.5$\,K) has been reported to become superconducting (SC) under $P$ \cite{Nakashima}. The most remarkable feature is that $P$-induced superconductivity emerges in two $P$ ranges of 1.7-3.7\,GPa and 5.9-7.3\,GPa, where zeroresistivity is observed \cite{Kotegawa}. Here, we denote the two SC domes in a low and high $P$ region as SC1 and SC2, respectively, as shown in Fig.~\ref{fig1}. In this figure, an application of $P$ makes $T_{\rm N}$ increase and exhibit its maximum at 8.5\,K around $P\sim 3$\,GPa, and disappears around $P\sim$ 7\,GPa \cite{Kotegawa,Harada}. Interestingly, SC1 seems to appear inside the AFM phase, exhibiting a maximum of $T_{\rm SC}\sim 0.3$\,K at the maximum value of $T_{\rm N}$, whereas SC2 emerges around the QCP where the AFM order collapses, as reported thus far \cite{Kotegawa}. The emergence of SC1 may be due to the delocalization of Ce-4$f$ electrons, even though the AFM order is robust against the application of $P$, which causes $T_{\rm N}$ and the internal magnetic field to increase \cite{Harada}. These experimental results suggest that the emergence of SC1 relates to a novel type of SC mechanism, which differs from that in SC2 near the QCP.

In this letter, we report on the novel characteristics of antiferromagnetism and SC1 observed from the $^{73}$Ge-NQR measurements under 
$P$ at zero field ($H=0$). We reveal that an unconventional SC1 emerges under the background of a commensurate AFM structure in which an incommensurate spin-density-wave order changes around $T\sim 2$\,K well below $T_{\rm N}$.  It is demonstrated that the $P$-induced SC1 in CeNiGe$_3$ occurs in such a characteristic AFM state that the application of $P$ increases $T_{\rm N}$ inside the AFM phase far from the QCP.

A polycrystalline sample enriched with $^{73}$Ge was crushed into a powder in order to allow RF pulses to penetrate into the samples easily for NQR measurements. The NQR measurements were performed by the conventional spin-echo method in a frequency ($f$) range of 2.8-10\,MHz at $P= 0$, 2.0, and 2.8\,GPa. A $^{3}$He-$^{4}$He dilution refrigerator was used to obtain the lowest temperature ($T$) of 50\,mK. A BeCu/NiCrAl piston-cylinder $P$ cell with a $P$-transmitting medium of polyethylsiloxane (PES-1) was used to generate hydrostatic $P$ \cite{Kirichenko}. The value of $P$ was determined from the $T_{\rm SC}$ of Sn obtained from the resistivity measurements.

Figure~2(a) indicates the well-articulated  spectra of $^{73}$Ge-NQR for a nuclear spin $I=9/2$ which arises from three Ge sites. By incorporating the NQR parameters ($\nu_{\rm Q}, \eta$) obtained from the band calculation by H.~Harima {\it et al.}, each NQR peak is considered to arise from the Ge1 ($V_{\rm zz}\parallel b$), Ge2 ($V_{\rm zz}\parallel c$), and Ge3 ($V_{\rm zz}\parallel b$) sites in CeNiGe$_3$ as shown in the figure where the simulated spectra are presented and each NQR frequency is indicated by an arrow \cite{Harada}. It should be noted that all these spectra are affected by the emergence of the  internal field $H_{\rm int}$ at each Ge site below $T_{\rm N}$, confirming absence of a paramagnetic (PM) phase at $P=0$, 2.0, and 2.8\,GPa. $T_1$ and $T_2$ were measured at the 4$\nu_{\rm Q}$($\pm7/2 \leftrightarrow \pm9/2$) transition of the Ge1 site  ($f\sim 6.3$\,MHz) because this transition is well resolved from the others below $T_{\rm N}$. On the other hand, in the PM at $P=0$, they were measured at the 2$\nu_{\rm Q}$($\pm3/2 \leftrightarrow \pm5/2$) transition of the Ge1 site ($f\sim 3.1$\,MHz) because the 4$\nu_{\rm Q}$ of the Ge1 site accidentally overlapped with the 3$\nu_{\rm Q}$ of the Ge3 site. The Lorenzian and Gaussian components $1/T_2$ were observed below and above $T_{\rm N}$, respectively. They were fitted with
 $M(2\tau)/M_0\propto\exp(-2\tau/T_2)$ (Lorenzian), and $M(2\tau)/M_0\propto\exp(-(2\tau)^2/2T_2^2)$ (Gaussian), respectively. Here, $\tau$ is the interval between two rf pulses ($\pi$/2 and $\pi$ pulses) in spin-echo method. 

We begin with the $P$-induced evolution of the antiferromagnetism in CeNiGe$_3$. The NQR spectra for the $4\nu_{\rm Q}$ transition of the Ge3 site in PM state at $P=0$, 2.0, and 2.8\,GPa are indicated in the upper parts of Fig.~2(b). 
As $T$ decreases below $T_{\rm N}$, the NQR spectrum exhibits a double-peak structure due to the appearance of $H_{\rm int}$ associated with the AFM order.  A separation between the two peaks in the NQR spectrum is related to the $H_{\rm int}$ that is proportional to the staggered magnetization $M_{\rm Q}$ below $T_{\rm N}$. By extrapolating the $T$ dependence of $H_{\rm int}$, we obtain $T_{\rm N}=5.1$, 7.4, and 8.2\,K and the saturated $H_{\rm int}=0.70$, 0.86, and 0.91\,kOe at $P=0$, 2.0, and 2.8\,GPa, respectively. As observed in the phase diagram in Fig.~1, the fact that $T_{\rm N}$ increases with an application of $P$ is corroborated by the present experiment. Since a hyperfine coupling constant $A_{\rm hf}$ is estimated to be $\sim$\,0.86\,kOe/$\mu_{\rm B}$ at the Ge3 site, by using $M_{\rm Q}\sim 0.8\mu_{\rm B}$ at ambient $P$ \cite{Durivault}, $M_{\rm Q}$ can be as large as  $\sim$\,1.0\,$\mu_{\rm B}$   
at 2.8\,GPa. These results indicate that the AFM order becomes robust against increasing values of $P$, where SC1 occurs with a maximum $T_{\rm SC}$ of 0.3\,K \cite{Kotegawa}.
\begin{figure}[h]
\centering
\includegraphics[width=7.1cm]{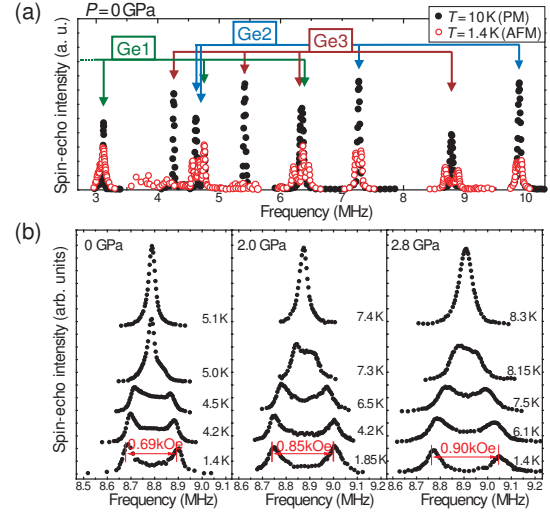}
\caption[]{(color online) (a) $^{73}$Ge-NQR spectra for the PM state at $T=10$\,K (solid circles) and for the AFM state at $1.4$\,K (open circles) at $P=0$\,GPa. 
 (b) Temperature dependence of the $4\nu_{\rm Q}$ transition for the Ge3 site ($f\sim8.8$\,MHz) at $P=0$, 2.0, and 2.8\,GPa. The spectra for the PM state are shown at the top of each figure, while the others for the AFM state at each $P$. }
\end{figure}

Figure~3 shows the $T$ dependence of 1/$T_1$ at the Ge1 site at $P=0$, 2.0, and 2.8\,GPa. In the PM state, the 1/$T_1$ at $P=0$ stays almost constant above $T_{\rm N}$, revealing that Ce-4$f$-derived moments behave as if they are localized. As $P$ increases, the $1/T_1$s at $P=2.0$ and 2.8\,GPa start to decrease below $T^*\sim 14$ and 18\,K, respectively, which are well above $T_{\rm N}\sim 8$\,K. It indicates the $P$-induced increase in hybridization with conduction electrons makes the 4$f$ electrons itinerant below $T^{\rm *}$, which resembles  the case of CeRhIn$_5$ \cite{Kawasaki2}. These results suggest that the $P$-induced increase in hybridization with conduction electrons may lead to an onset of SC1. However, $T_{\rm N}(P)$ and $H_{\rm int}(P)\propto M_Q(P)$ are not expected to increase in such an itinerant regime of Ce-4$f$ electrons.
\begin{figure}[h]
\centering
\includegraphics[width=7.2cm]{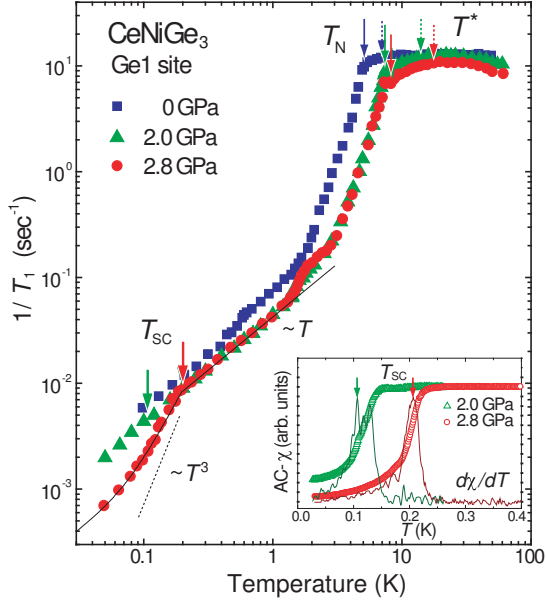}
\caption[]{(color online) Temperature dependence of $1/T_1$ at $P=0$, 2.0, and 2.8\,GPa at the Ge1 site. 
At $P=2.0$ and 2.8\,GPa, SC1 emerges inside the AFM phase far from the QCP. The solid curve below $T_{\rm SC}=0.2$\,K is obtained from a calculation based on a line node in the SC gap with $2\Delta_0/k_{\rm B}T_{\rm SC}=4.6$ and $N_{\rm res}/N_0=0.56$. The inset shows the $T$ dependences of ac-susceptibility (open symbols) and $d\chi_{\rm ac}/dT$ (dotted lines) at $P=2.0$ and 2.8\,GPa.}
\end{figure} 

Next, we consider the SC characteristics. As revealed by the $T$ dependence of $1/T_1$ at $P=2.8$\,GPa, an observation of $T_1T=$~const. behavior points to the presence of the Fermi surface in the AFM state well below $T_{\rm N}=8.2$\,K. Remarkably, a distinct decrease in $1/T_1T$  below $T_{\rm SC}=0.2$\,K provides microscopic evidence of the emergence of SC1, which coexists with the antiferromagnetism. In fact, this onset of SC1 at $P=2.8$\,GPa is corroborated by the diamagnetism in ac-susceptibility $\chi_{\rm ac}$ measured by the in-situ NQR coil,  as shown in the inset of Fig.~3. Here, the $T$-derivative of $\chi_{\rm ac}$, $d\chi_{\rm ac}/dT$, exhibits a peak at $T_{\rm SC}=0.2$\,K. It should be noted that a bulk $T_{\rm SC}=0.2$\,K is lower than $T_{\rm SC}\sim0.3$\,K obtained by resistivity at $P=2.9$\,GPa \cite{Kotegawa}. Below $T_{\rm SC}=0.2$\,K, $1/T_1$ shows no coherence peak, subsequently, it exhibits a power-law like $T^{\rm n}$ (n\,=\,2\,$\sim$\,3) variation, and upon further cooling, it approaches a $T$-linear behavior.  These behaviors are associated with an unconventional  superconductivity with a line node in the SC gap function with a finite residual density of state ($N_{\rm res}$) at the Fermi level, which  used to be observed in other HF SC compounds such as CeCu$_2$Si$_2$ \cite{Kitaoka}, CeRhIn$_5$ \cite{Kawasaki1,Yashima1}, and CeCoIn$_5$ \cite{Yashima2}. As shown by a solid curve in Fig.~3, the line-node gap model with $\Delta(\theta)=\Delta_0\cos\theta$ is consistent with the $1/T_1$ data below $T_{\rm SC}$ by assuming parameters of $2\Delta_0/k_{\rm B}T_{\rm SC}=4.6$ and $N_{\rm res}$/$N_0=0.56$. Here, $N_0$ is the density of state (DOS) at the Fermi level in the AFM state. 
The large fraction of $N_{\rm res}$/$N_0=0.56$ indicates a novel SC character with a finite weight of low-lying quasiparticle excitations due to the uniformly coexisting state of SC1 and AFM states as argued in CeRhIn$_5$ \cite{Kawasaki1,Yashima1}. As shown in the inset of Fig.~3, the $T$ dependences of ac-susceptibility and $d\chi_{\rm ac}/dT$ at $P=2.0$\,GPa reveal an onset of SC at $T_{\rm SC}\sim 0.11$\,K. We cannot observe a clear decrease in 1/$T_1$ below $T_{\rm SC}\sim 0.11$\,K; this suggests that a SC gap does not fully open at $P=2.0$\,GPa. A residual DOS is tentatively estimated to be as large as $N_{\rm res}$/$N_0\sim 0.95$. Evidently, some impurity effect fails to explain the $P$ dependence of the residual DOS such as $N_{\rm res}$/$N_0=0.95$ and 0.56 at $P=2.0$ and 2.8\,GPa, respectively.
\begin{figure}[hb]
\centering
\includegraphics[width=8.6cm]{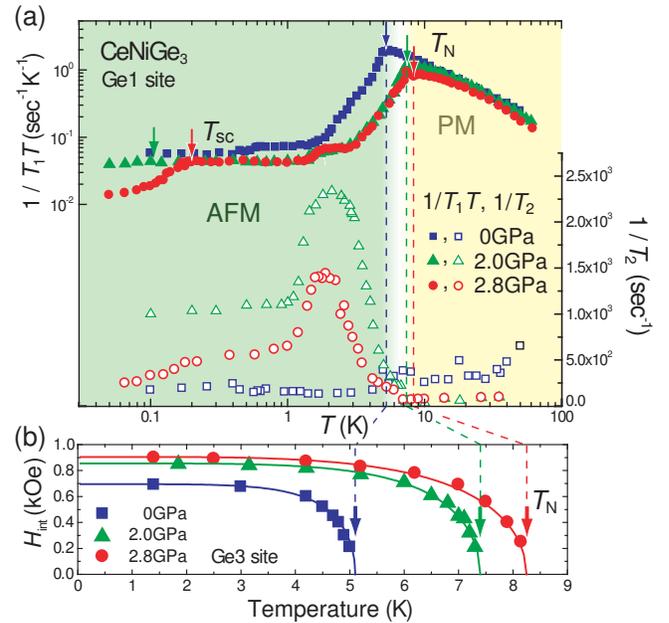}
\caption[]{(color online) (a) Temperature dependence of $1/T_1T$ and $1/T_2$ for the Ge1 site at $P=0$, 2.0, and 2.8\,GPa. Note that $1/T_2$ exhibits a distinct peak at $T\sim 2$\,K well below $T_{\rm N}$, but far above $T_{\rm SC}$; 1/$T_1T$ decreases at $P=2.0$ and 2.8\,GPa. (b) Temperature dependence of $H_{\rm int}(T)$ at $P=0$, 2.0, and 2.8\,GPa. The solid curves seen as a guide to the eye whose extrapolations to $T=0$ and $H_{\rm int}=0$ give rise to  $H_{\rm int}(0)=0.70$, 0.86, and 0.91\,kOe and $T_{\rm N}=5.1$, 7.4, and 8.2\,K at $P=0$, 2.0, and 2.8\,GPa, respectively.}
\end{figure} 

Figure~4 shows the $T$ dependence of the nuclear spin-spin relaxation rate $1/T_2$ at $P=0$, 2.0, and 2.8\,GPa at the Ge3 site, along with $1/T_1T$. It should be noted that $1/T_2$ can probe longitudinal low-lying magnetic excitations against the quantization axis of nuclear spin, whereas $1/T_1$ probes the transversal excitations. This is because $1/T_1$ and $1/T_2$ are generally given as $1/T_1 = A^2/2\hbar ^2 \int^{\infty}_{-\infty} \cos \omega_0 \tau~\langle [S_+(\tau )S_-(0)] \rangle ~d\tau$ and $1/T_2 = 1/2T_1 + A^2/2\hbar ^2 \int^{\infty}_{-\infty} \langle [S_z(\tau )S_z(0)] \rangle~d\tau$, where the $z$ axis is the quantization axis. Here, [$S_+(\tau )S_-(0)$] and [$S_z(\tau )S_z(0)$] correspond to  the $xy$-plane and the z components of the spin-spin correlation functions, respectively, and $A$ is the hyperfine coupling constant \cite{Slichter}. 
A remarkable result is that {\it $1/T_2$ gradually increases and exhibits a prominent peak at $T\sim 2$\,K in the AFM state} under $P$, whereas {\it $1/T_1T$ decreases suddenly at $T_{\rm N}$ and $T\sim 2$\,K}, as shown in Fig.~4. It should be noted that the peak in $1/T_2$ at $P=0$ is absent, but a distinct peak in $1/T_2$ appears on applying $P$. It is noteworthy that there is no report on such a marked peak in $1/T_2$ below $T_{\rm N}$ in other Ce-based superconductors. Since the anomaly at $T\sim 2$\,K has not been observed in specific-heat measurements under $P$ \cite{Tateiwa}, the anomaly cannot be attributed to some phase transition, but rather due to some changes in the characteristics of magnetic fluctuations or a gradual change of AFM structure. Furthermore, it is observed that the NQR spectral shape changes on cooling below $\sim$\,2\,K at $P=2.8$\,GPa, suggesting a  change of AFM structure. Therefore, we focus on the $T$-derived evolution of the NQR spectrum for the $4\nu_{\rm Q}$ transition at the Ge3 site. This is because the appearance of the largest value of $H_{\rm int}$ at the Ge3 site enables us to sensitively detect a possible change in the AFM structure. 
\begin{figure}[hb]
\centering
\includegraphics[width=8.5cm]{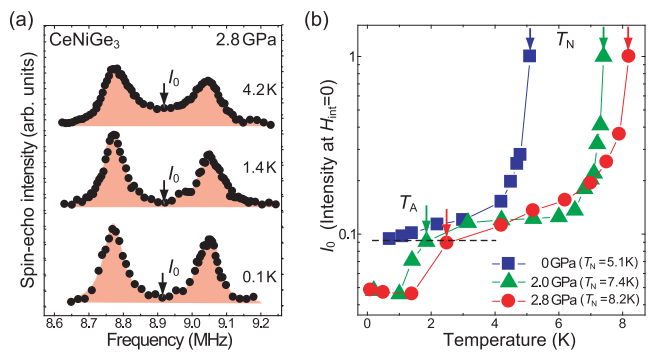}
\caption[]{(color online) (a) The NQR spectra for the AFM state at $T=4.2$, 1.4, and 0.1\,K lower than $T_{\rm SC}=0.2$\,K at $P=2.8$\,GPa. (b) Temperature dependence of the NQR intensity $I_{\rm 0}$ at the center with $H_{\rm int}=0$ at $P=0$, 2.0, and 2.8\,GPa. $I_{\rm 0}$ markedly decreases below $T_{\rm A}\sim 1.9$ and 2.5\,K at $P=2.0$ and 2.8\,GPa, respectively, suggesting the transition from the incommensurate spin-density-wave order into the commensurate AFM phase.}
\end{figure} 

In the AFM state at $P=0$\,GPa, the NQR intensity between the double peaks in the spectrum can be continuously observed, which suggests a possible distribution in $H_{\rm int}$ relevant to the incommensurate structure of the AFM order with a propagation vector $Q=$\,[0,\,0.409,\,0.5] by the neutron-powder diffraction at $P=0$ \cite{Durivault}.
At $P=2.0$ and 2.8\,GPa, however, it should be noted that as $T$ is reduced, the NQR intensity $I_0$ at the center with $H_{\rm int}=0$  decreases, and hence, each spectral width becomes sharp, as shown in Fig.~5(a). Indeed in Fig.~5(b), $I_{\rm 0}$ markedly decreases below $T_{\rm A}\sim 1.9$ and 2.5\,K at $P=2.0$ and 2.8\,GPa, respectively. This fact reveals that $M_{\rm Q}$ is no longer widely distributed at low $T$, suggesting the transition from the incommensurate spin-density-wave order to the commensurate AFM phase. These $T_{\rm A}$s coincide with $T\sim2$\,K around which $1/T_2$ exhibits the peak. Interestingly, as $T_{\rm A}$ and $T_{\rm N}$ increase with $P$, $T_{\rm SC}$ is enhanced as observed in Fig.~1. A commensurate AFM order should be stable, accompanying a change in the wave vector $Q_{\rm b}$ below $T_{\rm A}\sim 1.9$ and 2.5\,K at $P=2.0$ and 2.8\,GPa, respectively. This change in the AFM spin structure may spatially distribute the AFM spin density, causing longitudinal spin-density fluctuations on cooling. As a result, a longitudinal component of $H_{\rm int}$ fluctuates at the Ge1 site, which enables us to observe the peak in $1/T_2$. It is likely that the longitudinal spin-density fluctuations would be soften below $T_{\rm A}$, and they may be a mediator of SC1 emerging in the AFM state. Notably, this longitudinal spin-density fluctuation was also indicated at the end point of the quantum first-order transition from FM2 to FM1 in the ferromagnetic superconductor UGe$_2$ \cite{Huxley,Harada2}. We suggest that the longitudinal spin-density fluctuation may be the another possible mechanism for enhancing $T_{\rm sc}$ in magnetic state.

In conclusion, the $^{73}$Ge-NQR measurements of CeNiGe$_3$ under $P$ have revealed that the application of $P$ increases $T_{\rm N}$ and $T_{\rm A}$, around which a successive transition takes place from the incommensurate spin-density-wave order to the commensurate AFM. The unconventional SC1 emerges under the background of the commensurate antiferromagnetism far from the QCP, with an increase in $T_{\rm SC}$ as $T_{\rm N}$ and $T_{\rm A}$ also  increase.  A characteristic feature of SC1 is that a large fraction of low-lying excitations remains in the quasiparticle excitation spectrum along with the uniform coexistence with antiferromagnetism. These novel phenomena in CeNiGe$_3$ have hitherto never been reported in the $P$-$T$ phase diagrams among HF compounds. 
 
We would like to thank M.~Yashima, H.~Kotegawa, and T.~C.~Kobayashi for fruitful discussions and comments. This work was supported by a Grant-in-Aid for Creative Scientific Researchi15GS0213), the Ministry of Education, Culture, Sports, Science and Technology (MEXT) and the 21st Century COE Program (G18) supported by the Japan Society for the Promotion of Science (JSPS). A.H. was financially supported by a Grant-in-Aid for Exploratory Research of MEXT (No.~17654066).

\end{document}